# Electric field effect in ultrathin black phosphorus

Steven P. Koenig[1,2], Rostislav A. Doganov[1,2], Hennrik Schmidt[1,2], A. H. Castro Neto[1,2], and Barbaros Oezyilmaz[1,2]

[1]Graphene Research Centre, National University of Singapore, 6 Science Drive 2, Singapore 117546
[2]Department of Physics, National University of Singapore, 2 Science Drive 3, Singapore 117542

**Abstract:**

Black phosphorus exhibits a layered structure similar to graphene, allowing mechanical exfoliation of ultrathin single crystals. Here we demonstrate few-layer black phosphorus field effect devices on Si/SiO$_2$ and measure charge carrier mobility in a four-probe configuration as well as drain current modulation in a two-point configuration. We find room-temperature mobilities of up to 300 cm$^2$/Vs and drain current modulation of over $10^3$. At low temperatures the on-off ratio exceeds $10^5$ and the device exhibits both electron and hole conduction. Using atomic force microscopy we observe significant surface roughening of thin black phosphorus crystals over the course of 1 hour after exfoliation.

**Main Text:**

Two-dimensional crystals have gained significant attention since the rise in popularity of graphene[1]. Due to graphene's lack of a band gap the search for a two-dimensional gapped material with high mobility has been highly sought after. Other than graphene black phosphorus (BP) is the only other known monotypic van der Waals crystal[2–4]. Previous studies have shown that bulk BP has high carrier mobility combined with the presence of a direct band gap[5,6]. BP has also been shown to go through a series of phase transitions and becomes superconductive at high pressures[7–9]. The band gap of BP has been predicted to increase with decreasing number of layers from 0.3 eV in bulk to 2 eV for single layer[10–12]. This is similar to $MoS_2$, except that in BP the band gap remains direct for all number of layers[10]. Additionally, due to its relatively small band gap BP can be tuned into both p-type and n-type configuration in contrast to few-layer $MoS_2$ which usually exhibits n-type behavior[13].

The phosphorus atoms of BP covalently bond to three neighboring atoms but unlike graphene BP forms a puckered structure with out of plane ridges[14–16]. Individual layers of phosphorus atoms are held together by weak van der Waals forces[2,4]. Fig. 1(a) shows a schematic of the BP lattice. Unlike carbon, phosphorus has only three valance electrons which leads to BP being semiconducting since each atom is bonded to three neighboring atoms[10,17–19].

Few-layer BP was obtained by cleaving bulk BP crystals[20] using the well-known scotch-tape method. Exfoliated crystals were deposited onto $Si/SiO_2$ wafers with 300nm $SiO_2$, which is used as gate dielectric for our field effect devices. Thin crystals were identified using optical microscopy and further characterized with both AFM (atomic force microscope) and Raman spectroscopy. Fig. 1(b) shows an optical image of a typical thin flake and the insert shows the same flake with electrical contacts. The AFM characterization is depicted in Fig. 1(c). The Raman spectrum in the back-scattering configuration, which can be seen in Fig. 1(d), verifies the flake as thin BP. The $A_g^2$, $B_{2g}$, and $A_g^1$ peaks can be easily seen at wave numbers of 465.9, 438.3, and 359.6 cm$^{-1}$, respectively, which is in good agreement with previously reported results[21–24]. The $B_{1g}$ and $B_{3g}^1$ peaks are not detectable in Raman backscattering when the incident light is parallel to the b axis (perpendicular to the covalently bonded plane of P atoms) of the BP[25].

As can be seen from the AFM image in Fig. 1(c) there is significant roughness present on the exfoliated BP crystal. Despite being the most stable allotrope of phosphorus BP is not completely unreactive under ambient conditions with oxygen and water molecules[26,27]. The degradation of the surface with time was directly observed by measuring the roughness of a freshly exfoliated BP flake using AFM. RMS roughness versus time is plotted in Fig. 2. For this measurement a BP flake was exfoliated and continuously scanned using a 500nm x 500nm scan window. The first AFM image was taken approximately 21 min after the exfoliation (Fig. 2(a) first image) and the measured RMS roughness was 427 pm. After 30 min (Fig. 2(a) second image) the roughness increased to 589 pm and after 41 min (Fig. 2(a) third image) the roughness further increased to 977 pm. This rapid oxidation over a short period of time could contribute to both increased contact resistance to the BP and also lower carrier mobility.

In order to minimize the effect of crystal degradation, the exfoliated flakes which were used for device preparation where quickly covered with PMMA once being identified under optical microscopy. The average time to cover a flake with PMMA after exfoliation was less than 30 min. On selected flakes, electron-beam lithography was used to define electrical contacts. Ti/Au electrodes were fabricated by thermal evaporation of 5 nm Ti and 90 nm Au followed by standard lift-off in acetone. The samples were measured in both a four-point configuration using an AC lock-in amplifier and a two-point configuration using a DC signal. The four-point configuration was used to precisely extract the mobility while the two-point configuration was used to measure the drain current modulation.

Fig. 3(a) shows the four-point conductance versus back gate voltage taken at room temperature. A constant current of 50 nA was applied through the BP flake while $V_g$, the back gate voltage, was swept from -60 to +60 V. We extract the field effect mobility, $\mu_{FE}$, in the linear region of the conductance, $G$, using

$$\mu_{FE} = \frac{L}{W} \frac{1}{C_g} \frac{dG}{dV_g} \quad (1)$$

where $L$ and $W$ are the length and width of the channel respectively, and $C_g$ is the gate capacitance per unit area. A fit was taken in the linear region of the conductance versus back gate voltage measurement (Fig. 3(a), red dashed line) and using equation (1) a mobility of ~300 cm$^2$/Vs was extracted. The four-terminal configuration was used to avoid problems that arise with contact resistance in a two-terminal arrangement[28].

Figure 3(b) shows the *I-V* characteristics at room temperature at negative gate voltages. The linearity in the curves confirms that the contacts are ohmic at negative gate voltages where the device is in the metallic regime with holes as majority carriers. In the low conductance region the resistance of the device was comparable to the internal resistance of the lock-in amplifier so a two-point DC measurement was used to accurately measure the drain current modulation. Fig. 3(c) show the source drain current, $I_{sd}$, as a function of $V_g$ with a source-drain bias of 50 mV at temperatures from 5 K to 200 K. For T > 200K we observed a significant hysteresis depending on the gate sweep direction. Fig. 3(d) shows the hysteresis in our BP device measured at 250K with the inset showing the difference in threshold voltage[29] on the hole conduction side for the temperatures range studied. The positive direction of the hysteresis (as defined in ref. 30) suggests this behavior is due to charge trapping rather than capacitive coupling to the BP[30]. This is also supported by the observation that the hysteretic behavior disappears at low temperatures where trapping is suppressed (Fig. 3(d) inset)[30]. The fact that the hysteresis observed in BP is much higher than that observed in graphene and other 2D crystals further suggest that the surface degradation as observed in the AFM measurements could also contribute to the hysteretic behavior of BP[31,32]. At low temperatures (<200K) the hysteresis disappears, and the sample is ambipolar showing both electron and hole transport, depending on the polarity of the back gate voltage. Below 200K we measure a drain current modulation of over $10^5$, while the room-temperature value is $10^3$. The ambipolar transport shows an asymmetry between the hole and electron side. At negative gate voltages, on the hole side, the measured source-drain currents are much higher than the values on the electron side for equal doping levels. Furthermore, the increase of $I_{sd}$ with gate voltage is much steeper on the hole side, indicating a considerably higher mobility than on the electron side. The differences in transconductance characteristics on the electron and hole side reflect the band structure of few-layer BP. A higher mobility and lower effective mass on the hole side has been predicted theoretically for up to five BP layers[33]. The larger conductance at negative gate voltages could be due to a higher density of states on the hole side[10], although a difference in efficiency for hole and electron injection from the metal contacts to the BP could also play a role[34]. Another difference can be seen in the shift of threshold voltage for decreasing temperature. At low temperatures this shift is much larger for electrons, showing a significant difference in the low charge carrier density regime, where thermal activation and variable range hopping become important[35].

In conclusion, we showed the electric field effect of few-layer BP using two-point and four-point transport measurements. Drain current modulation and field effect mobility at room temperature were measured to be ~$10^3$ and ~300 cm$^2$/Vs, respectively. The mobility of few-layer BP measured here is lower than the 1,000 cm$^2$/Vs Hall mobility reported for bulk BP[36]. Using atomic force microscopy, the crystals were found to appreciably degrade in ambient conditions over the course of an hour after exfoliation. This degradation likely limits the performance of BP devices fabricated under ambient conditions. Fabrication in an environment with low oxygen content and low humidity will be needed to make devices with carrier mobilities comparable to the reported bulk values. For BP devices operated under ambient conditions encapsulation will be essential, for example using another thin material such as boron nitride. Upon overcoming these difficulties BP is a promising material for the next generation of two-dimensional electronic devices.

During the preparation of this manuscript we became aware of other works regarding electronic transport in black phosphorus, which have been published on a preprint server[37–39].

This work was supported by the Singapore National Research Foundation Fellowship award (RF2008-07-R-144-000-245-281), the NRF-CRP award (R-144-000-295-281), and the Singapore Millennium Foundation-NUS Research Horizons award (R-144-001-271-592; R-144-001-271-646). AHCN acknowledges the NRF-CRP award "Novel 2D materials with tailored properties: beyond graphene" (R-144-000-295-281)

**Figure Captions:**

FIG. 1. (a) Three dimensional lattice structure of monolayer black phosphorous. (b) Optical image of an exfoliated black phosphorus crystal on Si/SiO$_2$. The insert shows the same flake after fabrication of electrical contacts. The scale bar is 50 μm. (c) AFM image and (d) Raman spectrum of the exfoliated black phosphorus flake from (a) and (b)

FIG. 2. AFM images showing the surface of a thin black phosphorus flake (a) 21min, 30min and 41min (from left to right) after exfoliation. The scan window is 500nm x 500nm. All three images were taken at the same location on the surface of the flake. (b) Compilation of the measured RMS roughness (extracted from the AFM images) versus time after exfoliation.

FIG. 3. (a) Conductance versus back gate voltage in a four-point measurement configuration. The dashed line shows the slope of the linear part of the curve, from which a mobility of ~300 cm$^2$/Vs was extracted. (b) *I-V* characteristics in the two-point configuration at various gate voltages at room temperature for the same device as in (a). (c) Source-drain current versus gate voltage taken at various temperatures in a two-point configuration for the same device as in (a) and (b). (d) Source-drain current for positive and negative sweep directions at 250K with a sweep rate of 0.1 V/s. Inset shows the difference in the threshold voltage, ΔV, between the positive and negative gate sweep direction as a function of temperature.

**Figure 1**

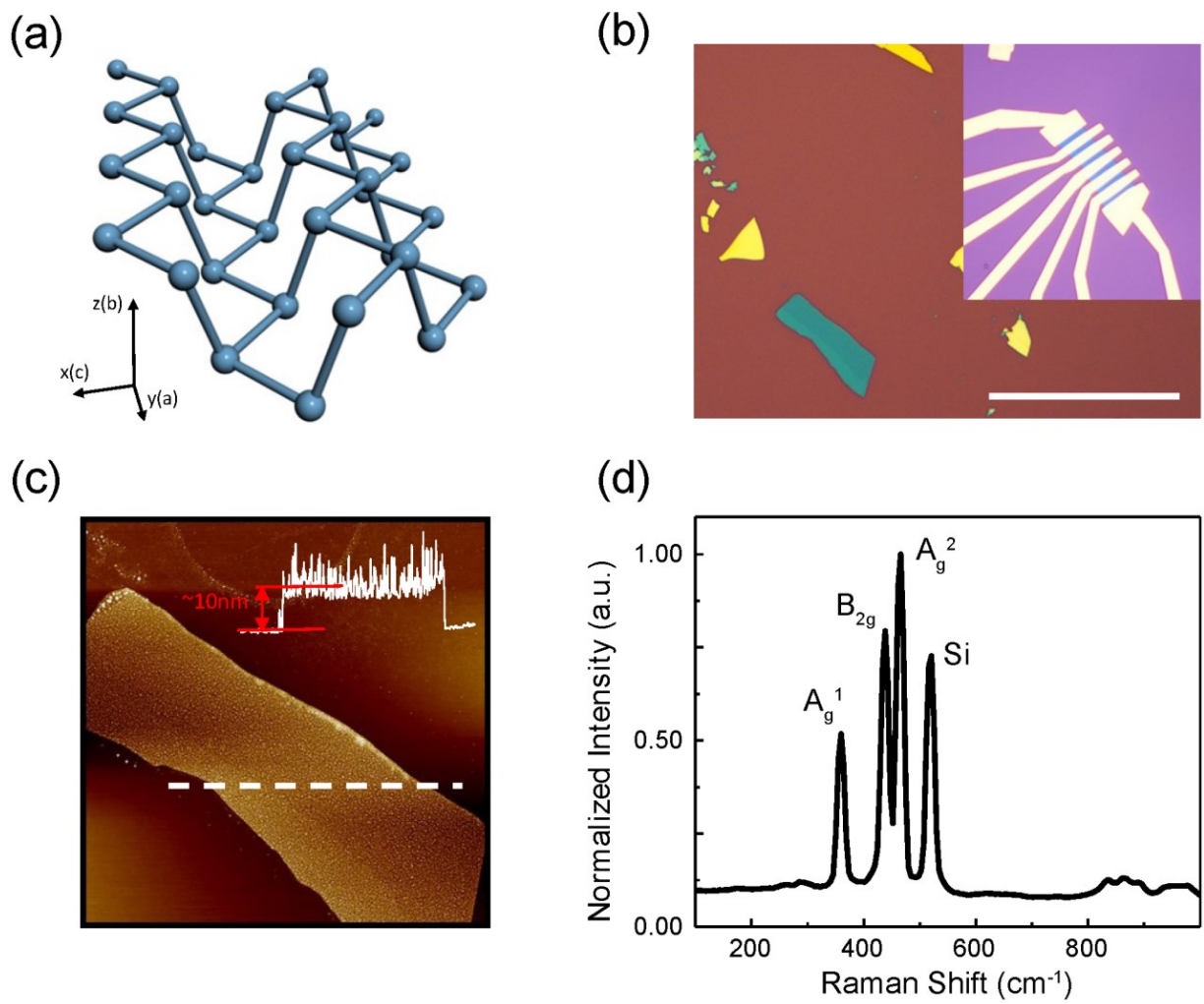

**Figure 2**

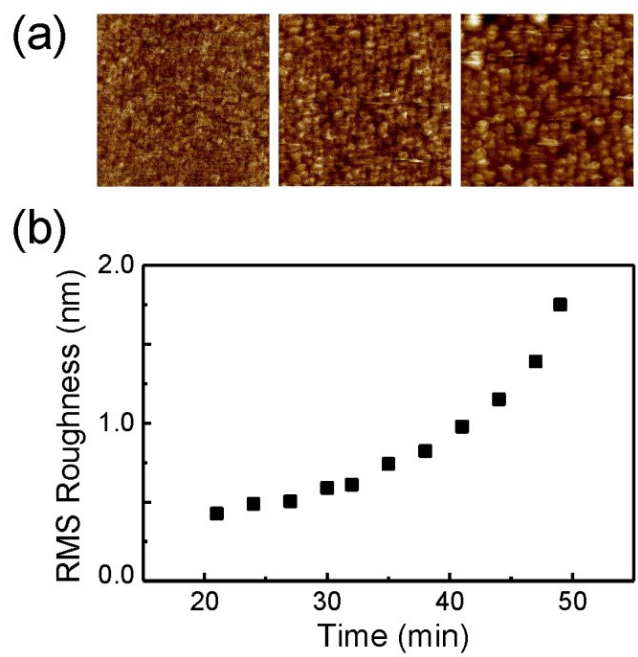

**Figure 3**

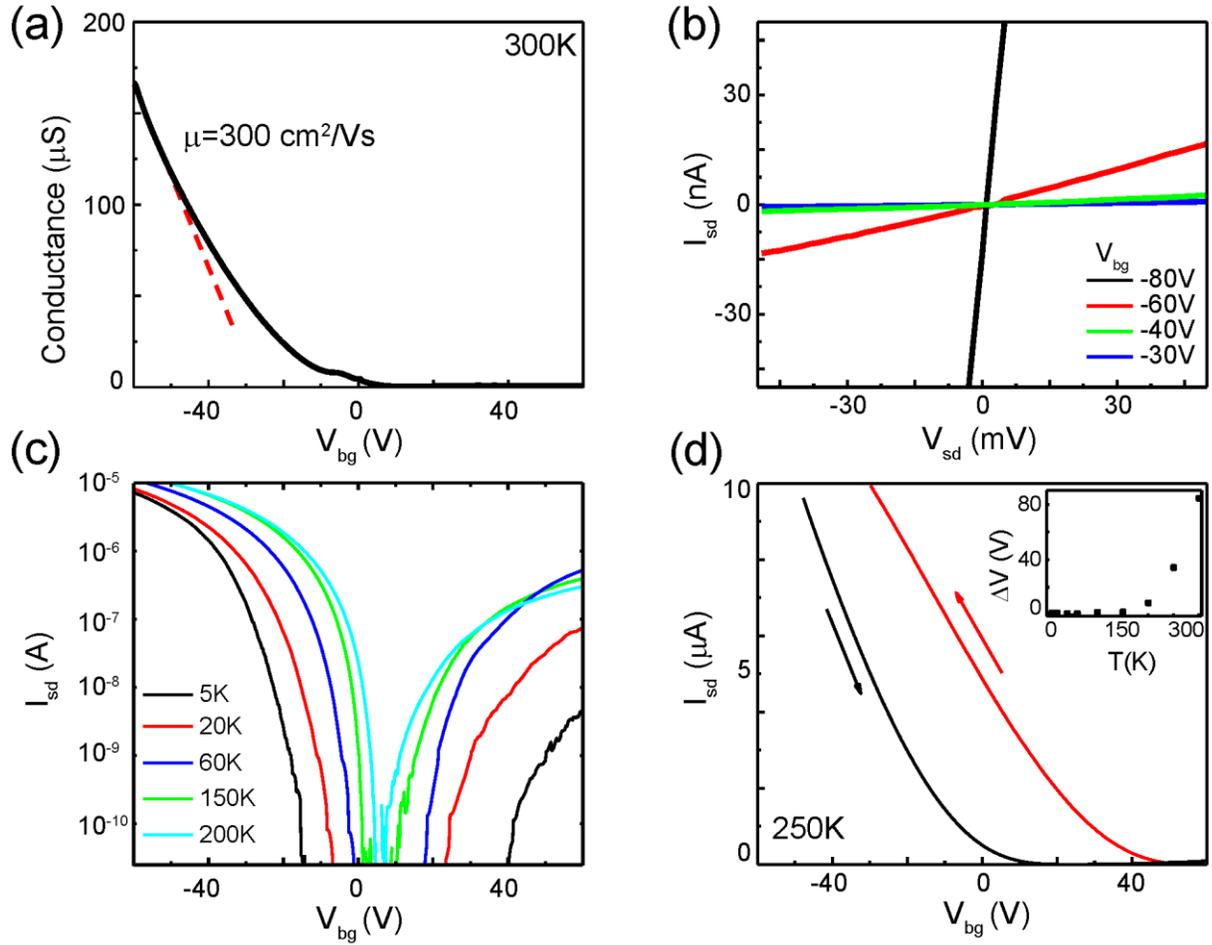